\documentclass[conf]{new-aiaa}
\usepackage[utf8]{inputenc}

\usepackage{graphicx}
\usepackage{amsmath}
\usepackage[version=4]{mhchem}
\usepackage{siunitx, tikz}
\usepackage{longtable,tabularx,array,floatrow}
\newfloatcommand{capbtabbox}{table}[][\FBwidth]
\usepackage{todonotes}
\usepackage{acro}
\DeclareAcronym{SCP}{
 short = SCP, 
 long = Sequential Convex Programming, 
}

\DeclareAcronym{6DOF}{
    short = 6-DoF,
    long = 6-degree of freedom,
}

\DeclareAcronym{5DOF}{
    short = 5-DoF,
    long = 5-degree of freedom,
}

\DeclareAcronym{RLV}{
    short = RLV,
    long = reusable launch vehicle,
}

\DeclareAcronym{ISA}{
    short = ISA,
    long = international standard atmosphere,
}

\DeclareAcronym{KSP}{
    short = KSP,
    long = Kerbal Space Program,
}

\DeclareAcronym{CTCS}{
    short = CTCS,
    long = continuous time constraint satisfaction,
}

\title{Ignition Point Reachability for Aerodynamically-Controlled Reusable Launch Vehicles}

\author[a]{Benjamin Chung~\footnote{Modeling \& Simulation Engineer, JuliaHub, Inc.\textit{Corresponding author email: }{\tt\footnotesize benjamin.chung@juliahub.com}.}}
\author[b]{Kazuya Echigo~\footnote{Doctoral Student, William E. Boeing Department of Aeronautics and Astronautics, University of Washington.}}
\author[b]{Beh\c{c}et A\c{c}{\i}kme\c{s}e~\footnote{AIAA Fellow. Professor, William E. Boeing Department of Aeronautics and Astronautics, University of Washington.}}
\affil[a]{JuliaHub, Inc}
\affil[b]{William E. Boeing Department of Aeronautics and Astronautics, University of Washington, Seattle, WA 98195, USA.\\}

\begin{document}

\maketitle

\begin{abstract}
We describe a successive convex programming (\ac{SCP}) based approach for estimate the set of points where a~\ac{5DOF}~\ac{RLV} returning to a landing site can transition from aerodynamic to propulsive descent. Determining the set of feasible ignition points that a~\ac{RLV} can use and then safely land is important for mission planning and range safety. However, past trajectory optimization approaches for~\acp{RLV} consider substantially simplified versions of the vehicle dynamics. Furthermore, prior reachability analysis methods either do not extend to the full constraint set needed for an~\ac{RLV} or are too beset by the curse of dimensionality to handle the full~\ac{5DOF} dynamics.

To solve this problem, we describe an algorithm that approximates the projection of a high dimensional reachable set onto a low dimensional space. Instead of computing all parts of the reachable space, we only calculate reachability in the projected space of interest by using repeated trajectory optimization to sample the reachable polytope in the reduced space. The optimization can take into account initial and terminal constraints as well as state and control constraints.

We show that our algorithm is able to compute the projection of a reachable set into a low dimensional space by calculating the feasible ignition points for a two-phase aerodynamic/propulsive RLV landing trajectory, while also demonstrating the aerodynamic divert enabled by our body and fin actuator model.
\end{abstract}
\acresetall

\section{Introduction}
Returning to land is an important part of the lifecycle of a reusable launch vehicle. Such return trajectories include both aerodynamic and propulsive descent guidance phases due to minimum throttle and relight count limitations. Most reusable launch vehicles have minimum thrust-to-weight ratio greater than 1 on touchdown, placing substantial pressure on trajectory generation.

As part of returning to a launch site a reusable vehicle must first reach a state from which it may propulsively land: too far away and it runs out of fuel before it touches down, too close and it hits the ground at substantial velocity. Moreover, range safety requirements frequently require that an aero-ballistic returning stage not be aimed at the landing site until it has successfully relit its engine and transitioned from aerodynamic to propulsive guidance~\cite{blackmore2016autonomous}. Consequently, a returning booster must maneuver aerodynamically to put its ignition point into a safe location and yet still be able to reach the desired landing site under power.

\begin{figure}[bt!]
\centering
\includegraphics[scale=0.8]{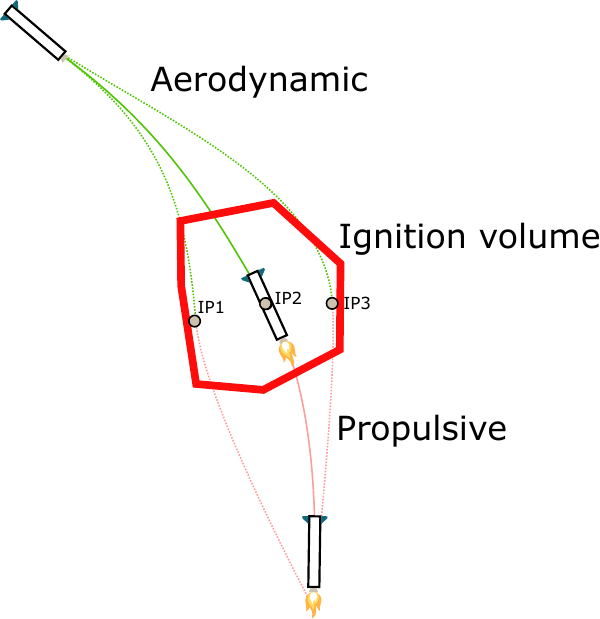}
\caption{Schematic representation of two-phase aerodynamic descent guidance.}
\label{fig:descent}
\end{figure}

The next question that arises is: given an initial state and a landing site where can the divert maneuver be performed? The returning stage can maneuver aerodynamically to reach the point from which it must be able to land without running out of fuel or shutting its engine down again as illustrated in Figure~\ref{fig:descent}. If we can compute such a reachable volume we can then choose where in the volume reignition occurs and powered descent guidance begins based on range safety and other concerns.

Trajectory optimization for low-fidelity~\ac{6DOF} reusable launch vehicle models has remained immature, with either simplistic drag-only aerodynamic models~\cite{szmuk2020successive} that are unsuitable for modeling aerodynamic maneuvers or reduced models with limited aerodynamic control modeling~\cite{doi:10.2514/6.2021-0862}. 

Reachability analysis adds a further challenge to the problem. Superficially, we wish to compute a 11-dimensional reachability analysis for a system subject to initial, final, and intermediate state and control constraints. This is untractable for most reachability techniques. For example, Hamilton-Jacobi methods are all beset by the curse of dimensionality~\cite{bansal2017hamilton} due to having to build and propagate representations of past reachable states while direct set computation approaches~\cite{althoff2021set} cannot guarantee satisfaction of future state constraints. Similarly, sampling-based methods~\cite{lew2021sampling} are limited in their ability to handle state and control constraints. 

Reachability of all dimensions and all time is intractable for our problem. However, we only care about a subset of the dimensions---the rocket's position---at a specific time: ignition. Accordingly, we are only truly interested in a 3-dimensional reachability problem for a single point. Reachability for our problem can therefore be reduced to identifying the volume in 3-dimensional space which is somehow reachable from the initial and final conditions in 11-dimensional space. We can do this using trajectory optimization.

We propose a trajectory optimization approach based on~\ac{SCP} for identifying the reachable set of ignition points for an aerodynamically-controlled launch vehicle. Previous work on trajectory optimization based reachability required a convex dynamics ~\cite{dueri2014finite} or only considered landing site reachability for the simpler exoatmopsheric lander dynamics~\cite{chan2022optimization}. By combining a new formulation of nonlinear axisymmetric body aerodynamics for better problem space exploration with trajectory optimization based sampling we can approximate reachable intermediate points even with highly nonlinear dynamics.

\section{Problem Description}

We aim to approximate the reachable volume for an intermediate point on a trajectory that satisfies initial and terminal conditions. In this section, we will describe the continuous-time nonlinear powered landing problem beginning with the vehicle dynamics, followed by the continuous-time objective, boundary conditions, and state and control constraints. 

\noindent We are focused on the two-phase powered descent guidance problem, where an aerodynamically and propulsively actuated rotationally axisymmetric vehicle undergoes two phases:
\begin{enumerate}
    \item Aerodynamic flight where the engine is shut off, and
    \item Propulsive flight after engine ignition.
\end{enumerate}
The time spent in each phase may vary so long as the boundary conditions and path constraints are satisfied. We wish to identify the volume in which the transition can occur while satisfying all other requirements.

\noindent Our model uses three reference frames:
\begin{enumerate}
    \item $F_I$, a surface fixed (rotating around the center of the planet) North-East-Up reference frame with the origin at the landing site.
    \item $F_B$, a vehicle-fixed frame centered at the vehicle center-of-mass, with its Z axis pointing out the nose of the vehicle and the X and Y axes chosen to be jointly perpendicular to Z and aligning with the axes in $F_I$ if the vehicle is unrotated.
    \item $F_W$, a vehicle-fixed frame centered at the vehicle center-of-mass, with its Z axis pointing opposite the air-relative velocity vector and the X and Y axes constructed from $F_I$ by the minimum rotation around $F_I$'s N/E axes such that the new Z axis is opposite the velocity vector.
\end{enumerate}
We notate the frame that a variable inhabits with a subscript; for example, $r_I(t)$ is the vehicle's position in the inertial frame. The construction of $F_B$ and $F_W$ will be covered in more detail later in this section.

\subsection{Vehicle Model}

\begin{figure}[hbt!]
    \centering
    \input{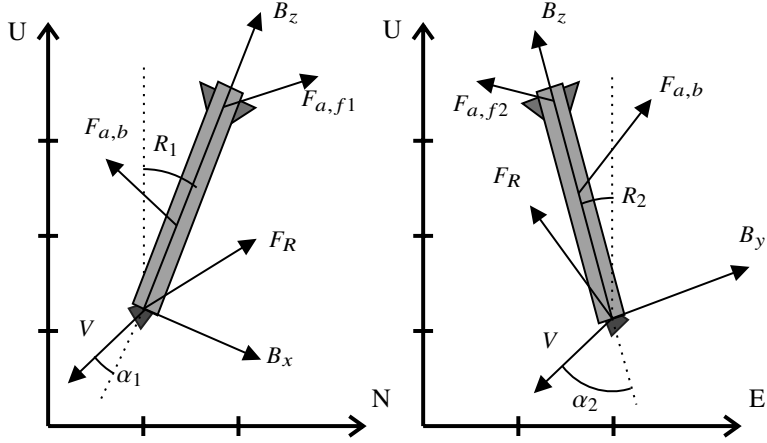}
    \caption{Vehicle dynamics}
    \label{fig:actuator}
\end{figure}

The vehicle is modeled as a rigid body. It possesses the states:
\begin{itemize}
    \item Mass: $m(t) \in \mathbb{R}$
    \item Position: $r_I(t) \in \mathbb{R}^3$
    \item Velocity: $v_I(t) \in \mathbb{R}^3$
    \item Body-to-inertial rotation: $R_{B/I}(t)  \in \mathbb{R}^2$
    \item Angular rate: $\omega_B(t) \in \mathbb{R}^2$
\end{itemize}
Since our vehicle is axisymmetric we only allow it to rotate along the north and east axes in the inertial frame; the rotation about the body up axis is fixed. Accordingly, our model is~\ac{5DOF} rather than~\ac{6DOF}. Consequently, we can use a simple pair of Euler angles to represent the vehicle's attitude. In order to reduce the number of needed discretization nodes we also apply a damping coefficient $C_\text{damp}$ to its attitude dynamics that approximates the affects of $C_{m,q}$ and $C_{m,\dot{\alpha}}$ though the same effect could be achieved with a feedback controller. 

\noindent Denoting the forces applied to the vehicle as $F_I(t)$ and torques $M_B(t)$, our basic equations of motion then follow as:
\begin{align}
\label{eqn:dyn:m} \dot{m}(t) &= \frac{||u_{B,R}(t)||_2}{g I_{sp}} \\
\label{eqn:dyn:r} \dot{r}_I(t) &= v_I(t) \\
\label{eqn:dyn:v} \dot{v}_I(t) &= \frac{F_I(t)}{m(t)} + A_{I,g}(t) + A_{I,c}(t) + A_{I,r}(t) \\
\label{eqn:dyn:f} F_I(t) &= T_{B/I}(R_{B/I}(t))(u_{B,R}(t)) + F_{I, fins}(t) + F_{I, body}(t) \\
\label{eqn:dyn:R} \dot{R}_{B/I}(t) &= \omega_B(t)\\ % should this be negated?
\label{eqn:dyn:om} \dot{\omega_B}(t) &= \begin{bmatrix} 1 & 0 & 0 \\ 0 & 1 & 0 \end{bmatrix} (J_B^{-1}(t) \times M_B(t)) - C_\text{damp} \omega_B(t) \\
\label{eqn:dyn:mb} M_B(t) &= r_{B, engine} \times u_B(t) + T_{I/B}(R_{B/I}) M_{I, body}(t) + M_{B, fins}(t) \\
\label{eqn:dyn:ijb} J_B(t) &= J_{B,dry} + \frac{m(t)-m_{dry}}{m_{wet}-m_{dry}}(J_{B,wet} - J_{B,dry})
\end{align}
where $\times$ is the cross product. We denote the transformation matrix from frame $B$ to $I$ as $T_{B/I}(R_{B/I}(t))$. 

\noindent Reusable launch vehicles tend to have a large fuel fraction relative to their structural mass and therefore their moment of inertia changes meaningfully with fuel depletion. We approximate the dependence with a linear interpolation between the wet and dry moments of inertia, as depicted in Eq.~\eqref{eqn:dyn:ijb}.

We have two types of control:
\begin{itemize}
    \item Aerodynamic, represented as a force $u_{W, a}(t)$ applied to the vehicle at a fixed position relative to its center of mass, and
    \item Propulsive, represented as a force $u_{B,R}(t)$ applied by the rocket at an offset $e_B$ from the vehicle's center of mass.
\end{itemize}
Propulsive control is dispatched with easily through inclusion in the mass depletion dynamics Eq.~\eqref{eqn:dyn:m}, in the translational dynamics Eq.~\ref{eqn:dyn:f} as transformed by the the body to inertial rotation, and by inclusion in the body rotational moment Eq.~\eqref{eqn:dyn:mb}. Aerodynamic controls will be discussed later.

\paragraph{Environmental Modeling}

We adopt the~\ac{ISA}~\cite{ISO2533} as our atmospheric model, which is used to determine the local atmospheric density $\rho(r_I)$, temperature, and speed of sound $c(r_I)$. Aerodynamic data is then normalized to the reference pressure $\rho_0$ which we define as sea level pressure and then scaled in the simulation by $\rho_r = \rho(r_I/\rho_0)$. Similarly, mach number is computed as $||v_I||_2/c(r_I)$. We assume that the surface-level temperature is unchanged from the~\ac{ISA} reference value.

\noindent We use a spherical gravity model $A_{I,g}(t) = \frac{\mu}{||r_I(t) - r_{I,ls}||_2^3}(r_I(t) - r_{I,ls})$ where $r_{I,ls}$ is the location of the center of the planet as expressed in the inertial reference frame. Furthermore, the planet (and thus our reference frame) are rotating with angular velocity $\omega_{I,p}$ inducing Coriolis $A_{I,c} = 2\omega_{I,p} \times v_I(t)$ and centrifugal accelerations $A_{I,r} = \omega_{I,p} \times (\omega_{I,p} \times (r_I(t) - r_{I,ls}))$.

\paragraph{Body Aerodynamics}
Aerodynamic modeling of axisymmetric bodies for optimization is challenging due to a singularity in the directions of lift and moment when the body is parallel to the velocity vector. Wings or other lifting devices would establish directions from which we define lift and slip axes but axisymmetric bodies offer no such affordances. 

\noindent We define drag to lie parallel to the flight path vector and lift to be pointed away from the flight path in the direction that the vehicle is tilted, thereby setting sideslip to be zero. Non-axisymmetric vehicles need to roll to maintain an explicit zero-sideslip orientation~\cite{dukeman2003enhancements} as their rotation around their forward vector changes the applied aerodynamic forces. However, our vehicle has no aerodynamic dependence on roll angle since it is axisymmetric.

\noindent The aerodynamic moment on a cylinder will be applied on an axis perpendicular to the body up and flight path vectors or $\hat{d}_{I,m}(t) = \frac{1}{||v_I(t)||_2\sin(\alpha(t))}\hat{b}_{I,z}(t) \times v_I(t)$. Lift is applied perpendicular to the flight path vector and the aerodynamic moment vector or $\hat{d}_{I, L}(t) = \frac{1}{||v_I(t)||_2}v_I(t) \times \hat{d}_{I,m}(t)$. The flight path direction $\hat{v}_I(t)$, the moment direction $\hat{d}_{I,M}$, and the lift direction $\hat{d}_{I,L}$ now forms an orthogonal basis.

\noindent Accordingly, basic equations for aerodynamic lift, drag, and moment are
\begin{align}
L_I(t) &= \frac{1}{2}\ \rho(t)\ C_l(\alpha(t),M(t))\ ||v_I(t)||_2^2\ \hat{d}_{I, L}(t) \\
D_I(t) &= -\frac{1}{2}\ \rho(t)\ C_d(\alpha(t),M(t))\ ||v_I(t)||_2\ v_I(t)\\
M_I(t) &= \frac{1}{2}\ \rho(t)\ C_m(\alpha(t),M(t))\ ||v_I(t)||_2^2\ \hat{d}_{I, m}(t) \\
F_{I, body}(t) &= L_I(t) + D_I(t).
\end{align}
This version of drag has no singularities and can thus readily be used for optimization. Observe, however, that the computation of $\hat{d}_{I,m}(t)$ normalizes the cross product $\hat{b}_{I,z} \times v_I(t)$ by dividing by the angle of attack $\alpha(t)$. Accordingly, a singularity exists at 0 degrees angle of attack when there is no tilt angle to reference our frame to. This is undesirable as the reusable launch vehicle must land at a very low angle of attack. 

\noindent We simplify the lift formulation to solve this problem with two observations:
\begin{enumerate}
    \item At $\alpha(t) = 0$ the value of $C_l(\alpha(t), M(t))$ is not relevant since the body-up vector $\hat{b}_{I,z}$ is parallel to $v_I(t)$ and thus $||v_I(t) \times (\hat{b}_{I,z}(t) \times v_I(t))||_2 = 0$. 
    \item Lift of axisymmetric bodies typically has a nearly linear relationship to $\alpha$ at low angles of attack.
\end{enumerate}
These properties can be exploited to smooth out the singularity by factoring $\frac{1}{\sin(\alpha)} = \csc(\alpha)$ into the lookup table for $C_l$ to form the lookup table $C_{l,\text{mod}}$.
\begin{equation}
C_{l,\text{mod}}(\alpha, M) = \left\{\begin{array}{ll}C_{l}(\alpha, M) \csc(\alpha) & \alpha \neq 0 \\ 
0 & \alpha = 0 \end{array}\right.
\end{equation}
For nonzero angles of attack, we premultiply the value in the lookup table by $\csc(\alpha)$, while at 0 angle of attack we simply let the lookup table value be 0. The linearity of lift and torque at small $\alpha$ causes error to remain small between 0 and the first nonzero data point, while no lift or torque is produced at 0 angle of attack. 

\noindent Including the normalization $\csc(\alpha)$ in the lookup table then lets us define body lift and moment as follows:
\begin{align}
L_I(t) &= \frac{1}{2}\ \rho(t)\ C_{l,mod}(\alpha(t),M(t))\ v_I(t) \times (\hat{b}_{I,z}(t) \times v_I(t))\\
M_{I, body} &= \frac{1}{2}\ \rho(t)\ C_{m,mod}(\alpha(t),M(t))\ ||v_I(t)||_2\ (\hat{b}_{I,z}(t) \times v_I(t)).
\end{align}
The velocity normalization factors $||v_I(t)||_2^2$ are factored singly and doubly into the cross products for torque and lift, respectively, and including $\csc(\alpha(t))$ in the lookup table $C_{m,mod}(\alpha(t), m)$ allows us to use the cross product magnitude to calculate lift and drag directly.

\begin{figure}[hbt!]
\begin{floatrow}
\ffigbox{%
\includegraphics[scale=0.7]{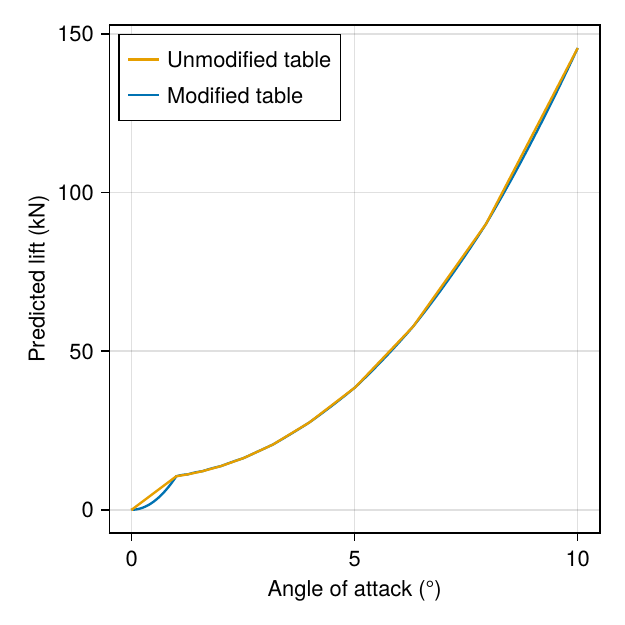}
}{
\caption{Modified vs. unmodified lift predictions at mach 0.9.}
\label{fig:approx-abserror}
}
\ffigbox{%
\includegraphics[scale=0.7]{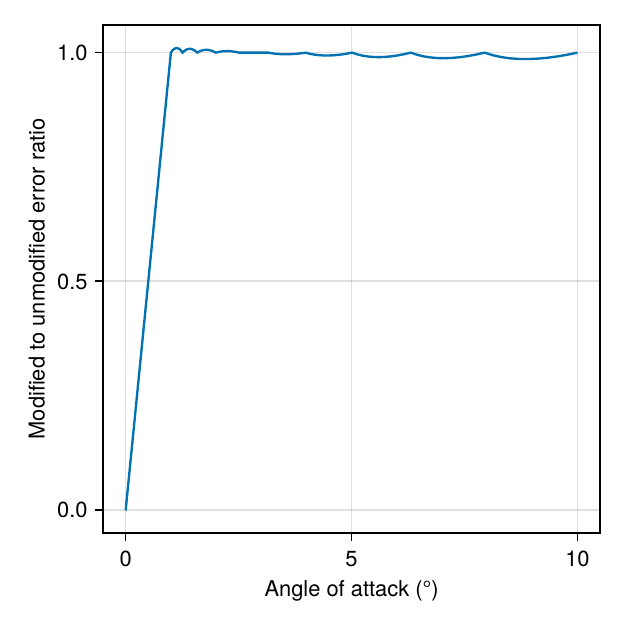}
}{
\caption{Relative lift force error vs. original lookup table at mach 0.9.}
\label{fig:approx-relerror}
}
\end{floatrow}
\end{figure}

Figures~\ref{fig:approx-abserror} and~\ref{fig:approx-relerror} show the relative and absolute force error as a function of angle of attack at mach 0.9. We evaluated the reference medium-fidelity lift and drag models for the vehicle in a fixed 0 degree fin actuation down to \qty{1}{\deg} angle of attack. Out to the limit of the underlying look up table, the modified lookup table maintains good agreement ($< 2 \%$) to the linear interpolation of the original look up table. However, below the minimum available angle of attack the approximation loses validity. Absolute error below the end of the lookup table is small in absolute terms and proportional to the force at the cutoff minimum nonzero angle of attack in the reference lookup table.

\paragraph{Actuator Aerodynamics}

\begin{figure}[hbt!]
\begin{floatrow}
\ffigbox{%
\tikzset{every picture/.style={line width=0.75pt}} %set default line width to 0.75pt        

\begin{tikzpicture}[x=0.75pt,y=0.75pt,yscale=-0.8,xscale=0.8]
%uncomment if require: \path (0,421); %set diagram left start at 0, and has height of 421

%Shape: Axis 2D [id:dp6443875700766978] 
\draw  (98.87,296.97) -- (323.1,296.97)(210.43,199.1) -- (210.43,391.8) (316.1,291.97) -- (323.1,296.97) -- (316.1,301.97) (205.43,206.1) -- (210.43,199.1) -- (215.43,206.1) (284.43,291.97) -- (284.43,301.97)(136.43,291.97) -- (136.43,301.97)(205.43,222.97) -- (215.43,222.97)(205.43,370.97) -- (215.43,370.97) ;
\draw   (291.43,308.97) node[anchor=east, scale=0.75]{1} (143.43,308.97) node[anchor=east, scale=0.75]{-1} (207.43,222.97) node[anchor=east, scale=0.75]{1} (207.43,370.97) node[anchor=east, scale=0.75]{-1} ;
%Shape: Rectangle [id:dp20439833931300244] 
\draw  [fill={rgb, 255:red, 218; green, 218; blue, 218 }  ,fill opacity=0.4 ] (166.87,238.1) -- (271.1,238.1) -- (271.1,358.8) -- (166.87,358.8) -- cycle ;
%Shape: Rectangle [id:dp8661839783248148] 
\draw  [color={rgb, 255:red, 179; green, 179; blue, 179 }  ,draw opacity=1 ][dash pattern={on 0.84pt off 2.51pt}] (135.43,222.97) -- (283.43,222.97) -- (283.43,370.97) -- (135.43,370.97) -- cycle ;
%Straight Lines [id:da19855923899377192] 
\draw [color={rgb, 255:red, 135; green, 135; blue, 135 }  ,draw opacity=1 ]   (129.85,270.08) -- (166.85,283.08) ;
%Straight Lines [id:da24441838581783282] 
\draw [color={rgb, 255:red, 135; green, 135; blue, 135 }  ,draw opacity=1 ]   (291.83,211.33) -- (250.85,238.08) ;
%Straight Lines [id:da44140290260316273] 
\draw [color={rgb, 255:red, 135; green, 135; blue, 135 }  ,draw opacity=1 ]   (249.85,359.08) -- (250.85,385.08) ;
%Straight Lines [id:da18083385979956312] 
\draw [color={rgb, 255:red, 135; green, 135; blue, 135 }  ,draw opacity=1 ]   (270.85,328) -- (304.85,328.08) ;
%Straight Lines [id:da8981806335253083] 
\draw    (166.87,122.85) -- (166.87,238.1) ;
%Straight Lines [id:da8908231907540649] 
\draw    (271.1,89.48) -- (271.1,238.1) ;
%Shape: Axis 2D [id:dp7296518857041684] 
\draw [color={rgb, 255:red, 0; green, 0; blue, 0 }  ,draw opacity=1 ] (97.87,157.67) -- (322.1,157.67)(210.1,60.1) -- (210.1,157.67) (315.1,152.67) -- (322.1,157.67) -- (315.1,162.67) (205.1,67.1) -- (210.1,60.1) -- (215.1,67.1) (284.1,152.67) -- (284.1,162.67)(136.1,152.67) -- (136.1,162.67)(205.1,83.67) -- (215.1,83.67) ;
\draw   ;
%Curve Lines [id:da3915614270778637] 
\draw [color={rgb, 255:red, 74; green, 144; blue, 226 }  ,draw opacity=1 ][line width=1.5]    (142.1,70.67) .. controls (169.1,136.67) and (218.1,226.67) .. (279.1,69.67) ;

% Text Node
\draw (328,287) node [anchor=north west][inner sep=0.75pt]   [align=left] {$\displaystyle C_{l,1}$};
% Text Node
\draw (177,374) node [anchor=north west][inner sep=0.75pt]   [align=left] {$\displaystyle C_{l,2}$};
% Text Node
\draw (292,197) node [anchor=north west][inner sep=0.75pt]   [align=left] {$\displaystyle C_{l,2+}( M,\alpha _{1} ,\alpha _{2})$};
% Text Node
\draw (221,377) node [anchor=north west][inner sep=0.75pt]   [align=left] {$\displaystyle C_{l,2-}( M,\alpha _{1} ,\alpha _{2})$};
% Text Node
\draw (18,257) node [anchor=north west][inner sep=0.75pt]   [align=left] {$\displaystyle C_{l,1-}( M,\alpha _{1} ,\alpha _{2})$};
% Text Node
\draw (306,320) node [anchor=north west][inner sep=0.75pt]   [align=left] {$\displaystyle C_{l,1+}(M,\alpha _{1} ,\alpha _{2})$};
% Text Node
\draw (195,36) node [anchor=north west][inner sep=0.75pt]   [align=left] {$\displaystyle C_{d,1}$};
\end{tikzpicture}
}{
\caption{Fin control constraints and drag polar coupling.}
\label{fig:finpolar}
}
\ffigbox{%
\tikzset{every picture/.style={line width=0.75pt}} %set default line width to 0.75pt        

\begin{tikzpicture}[x=0.75pt,y=0.75pt,yscale=-1,xscale=1]
%uncomment if require: \path (0,399); %set diagram left start at 0, and has height of 399

%Straight Lines [id:da6035771676391399] 
\draw    (139.54,185.97) -- (139.63,262.97) ;
\draw [shift={(139.63,265.97)}, rotate = 269.93] [fill={rgb, 255:red, 0; green, 0; blue, 0 }  ][line width=0.08]  [draw opacity=0] (8.93,-4.29) -- (0,0) -- (8.93,4.29) -- cycle    ;
%Straight Lines [id:da6257300913001592] 
\draw    (139.54,185.97) -- (139.49,147.97) ;
\draw [shift={(139.49,144.97)}, rotate = 89.93] [fill={rgb, 255:red, 0; green, 0; blue, 0 }  ][line width=0.08]  [draw opacity=0] (8.93,-4.29) -- (0,0) -- (8.93,4.29) -- cycle    ;
%Straight Lines [id:da1069299063339667] 
\draw    (139.54,185.97) -- (187.54,185.91) ;
\draw [shift={(190.54,185.91)}, rotate = 179.93] [fill={rgb, 255:red, 0; green, 0; blue, 0 }  ][line width=0.08]  [draw opacity=0] (8.93,-4.29) -- (0,0) -- (8.93,4.29) -- cycle    ;
%Straight Lines [id:da6079143307652229] 
\draw [color={rgb, 255:red, 198; green, 198; blue, 198 }  ,draw opacity=1 ] [dash pattern={on 0.84pt off 2.51pt}]  (139.49,144.97) -- (190.49,144.77) ;
%Straight Lines [id:da9091678468410687] 
\draw [color={rgb, 255:red, 198; green, 198; blue, 198 }  ,draw opacity=1 ] [dash pattern={on 0.84pt off 2.51pt}]  (190.54,185.91) -- (190.5,153.77) -- (190.49,144.77) ;
%Straight Lines [id:da8545225154182902] 
\draw    (139.54,185.97) -- (188.15,146.66) ;
\draw [shift={(190.49,144.77)}, rotate = 141.04] [fill={rgb, 255:red, 0; green, 0; blue, 0 }  ][line width=0.08]  [draw opacity=0] (8.93,-4.29) -- (0,0) -- (8.93,4.29) -- cycle    ;
%Straight Lines [id:da3102005358945943] 
\draw    (139.54,185.97) -- (83.73,120.67) ;
\draw [shift={(81.78,118.39)}, rotate = 49.48] [fill={rgb, 255:red, 0; green, 0; blue, 0 }  ][line width=0.08]  [draw opacity=0] (8.93,-4.29) -- (0,0) -- (8.93,4.29) -- cycle    ;
%Shape: Arc [id:dp5991614712237741] 
\draw  [draw opacity=0] (158.99,208.33) .. controls (153.66,212.98) and (146.94,215.46) .. (140.12,215.6) -- (139.54,185.97) -- cycle ; \draw   (158.99,208.33) .. controls (153.66,212.98) and (146.94,215.46) .. (140.12,215.6) ;  
%Straight Lines [id:da6239637028812896] 
\draw  [dash pattern={on 0.84pt off 2.51pt}]  (139.54,185.97) -- (177.82,229.28) ;
%Shape: Axis 2D [id:dp6960615189631001] 
\draw [color={rgb, 255:red, 0; green, 0; blue, 0 }  ,draw opacity=1 ][line width=1.5]  (64.25,294.41) -- (210.82,294.41)(64.25,92.15) -- (64.25,294.41) -- cycle (203.82,289.41) -- (210.82,294.41) -- (203.82,299.41) (59.25,99.15) -- (64.25,92.15) -- (69.25,99.15) (112.25,289.41) -- (112.25,299.41)(160.25,289.41) -- (160.25,299.41)(59.25,246.41) -- (69.25,246.41)(59.25,198.41) -- (69.25,198.41)(59.25,150.41) -- (69.25,150.41) ;
\draw   ;

% Text Node
\draw (51.81,73.22) node [anchor=north west][inner sep=0.75pt]  [font=\small] [align=left] {$\displaystyle -V$};
% Text Node
\draw (133.47,127.31) node [anchor=north west][inner sep=0.75pt]  [font=\small] [align=left] {$\displaystyle F_{d,i}$};
% Text Node
\draw (192.47,178.72) node [anchor=north west][inner sep=0.75pt]  [font=\small,rotate=-358.89] [align=left] {$\displaystyle F_{l,i}$};
% Text Node
\draw (188.91,125.7) node [anchor=north west][inner sep=0.75pt]  [font=\small] [align=left] {$\displaystyle F_{a,f,i}$};
% Text Node
\draw (76.04,100.83) node [anchor=north west][inner sep=0.75pt]  [font=\small] [align=left] {$\displaystyle B_{z}$};
% Text Node
\draw (143.03,217.72) node [anchor=north west][inner sep=0.75pt]  [rotate=-339.8] [align=left] {$\displaystyle \alpha_{i}$};
% Text Node
\draw (132.81,266.22) node [anchor=north west][inner sep=0.75pt]  [font=\small] [align=left] {$\displaystyle V$};
% Text Node
\draw (216.81,285.22) node [anchor=north west][inner sep=0.75pt]  [font=\small] [align=left] {$ $};
% Text Node
\draw (216.81,285.22) node [anchor=north west][inner sep=0.75pt]  [font=\small] [align=left] {$\displaystyle L_{1}$};

\end{tikzpicture}
}{
\caption{Fin force frame.}
\label{fig:fin-force-frame}
}
\end{floatrow}
\end{figure}

Most reusable launch vehicles use four fins in a cross pattern at the top of the first stage for aerodynamic control. For simplicity, we consider a vehicle with fins located \qty{90}{\deg} around the circumference of the vehicle; skewed configurations can be analyzed with the same framework but with different force vectors, as discussed later.

Ultimately, we wish to abstract over exactly how fins produce lift and drag. To reduce the dimensionality of the control vector, the aerodynamic control vector is expressed as a force applied to the vehicle on a plane perpendicular to the velocity vector, and is thus two dimensional. This lift produced by the fins then incurs an induced drag on the vehicle. Furthermore, for any given attitude and mach number the fins can only produce so much lift before they begin stalling.

Our aerodynamic actuators apply a force calculated in the wind frame; our control is a vector in the plane perpendicular to the velocity vector that then induces a force opposite the velocity vector (drag). The geometry in the plane of a single fin pair is shown in Figure~\ref{fig:fin-force-frame}. Aerodynamic lift is scheduled through a table-derived scale factor from a normalized -1 to 1 input on each axis. The scale factor is determined by lookup table as a function of the current mach number and per-plane angles of attack $C_{fin,l,i}(M(t),\alpha_1(t),\alpha_2(t))$ that is normalized by the squared velocity. The angles of

To account for state-dependent lift control asymmetries like actuator rotation limits and critical angles of attack we apply an additional table-derived nonlinear box constraint on the control as shown in Figure~\ref{fig:finpolar}. For each condition the normalized control $u_a$ must satisfy the constraint
\begin{equation}
    \label{cst:aerocontrol} C_{l,i-}(M,\alpha _{1} ,\alpha _{2}) \leq u_{a,i} \leq C_{l,i+}(M,\alpha _{1} ,\alpha _{2}) \text{ for } i\in 1,2.
\end{equation} 

Induced drag is then coupled to lift through a table-derived drag polar as a function of mach. We observe that for a given mach and AoA pair induced drag is proportional to squared lift plus a sweep offset that is linear in the cosine of the opposing angle of attack and the lift squared.

Accordingly, our raw aerodynamic control dynamics in the wind frame are
\begin{align}
    f_{f,l,i}(t) &= u_{a,i}(t) ||v(t)||_2^2  C_{fin,l,i}(M(t),\alpha_1(t),\alpha_2(t)) \\
    f_{f,d}(t) &= C_{fin,d,lin}(M(t))\begin{bmatrix}\cos(\alpha_2(t)) & \cos(\alpha_1(t))\end{bmatrix} \begin{bmatrix}f_{f,l,1}^2(t) \\ f_{f,l,2}^2(t)\end{bmatrix} + C_{fin,d,cst}(M(t)) f_{f,l}^T(t) f_{f,l}(t) \\
    F_{W,f}(t) &= \begin{bmatrix} f_{f,l,1}(t) \\ f_{f,l,2}(t) \\ f_{f,d}(t) \end{bmatrix}.
\end{align}

The actuator forces must then be mapped from the wind frame to the inertial frame. The drag vector is easy: it merely points reverse of the velocity vector. The lift vectors are harder, as their clock angle around the velocity vector is determined by the vehicle roll angle. In our problem we fix the vehicle roll angle to align the body axes with the axes of the inertial frame, so the actuator lift basis vectors are a function only of the velocity. 

We orient the lift vectors by constructing a reference frame whose Z axis is aligned with the velocity vector. Our construction takes the unrotated body basis vectors (which are initially expressed in the NED inertial reference frame) and then rotates them so that the Z axis points through the velocity vector. We construct the rotation matrix $R_{W/I}$ using Möller and Hughes's construction~\cite{doi:10.1080/10867651.1999.10487509} that rotates $-\hat{Z}$ around the $-Z\times \hat{v}$ axis by the smallest angle such that $-R_{W/I} \hat{Z}$ is aligned with $\hat{v}$. This construction avoids any normalization beyond the computation of $\hat{v}$, thereby limiting the singularities to those that occur at zero velocity and at 180 degrees angle of attack. $R_{W/I}$ then relates the forces calculated in the unrotated inertial axis aligned body frame to the forces applied in the inertial frame from the wind frame.

The moment from the actuators is calculated trivially as the cross product of the force and the (assumed to be fixed) mean position of the actuators relative to the body center of mass. The net body-frame contribution of the aerodynamic actuators to force and moment is therefore:
\begin{align}
F_{I, fins}(t) &= R_{W/I}(t) f_{f,l}(t) - v_I(t) * f_{f,d}(t)\\
M_{B, fins}(t) &= \sum_i r_{B,fins} \times R_{I/B}(t) F_{I, fins}(t) .
\end{align}
Where $r_{B,fins}$ is the mean position of the fins in the body frame.

\subsection{Dynamic constraints}

\noindent We enforce initial and final position, velocity, orientation, and angular rate constraints:
\begin{align}
     \label{cst:ip} r_I(0)= r_{I,i},\ v_I(0)= v_{I,i},\ R_{B/I}(0)= R_{B/I, i},\ \omega_B(0)= \omega_{B,i},\ m(0)=m_i  \\
     \label{cst:fp} r_I(t_f)= r_{I,f},\ v_I(t_f)= v_{I,f},\ R_{B/I}(t_f)= R_{B/I, f},\ \omega_B(t_f)= \omega_{B,f} 
\end{align}

\noindent We similarly enforce a mass lower bound:
\begin{equation}
    \label{cst:mdry} m(t) \geq m_{dry} 
\end{equation}

\noindent The thrust vector is constrained to be within the gimbal range of the engine $\theta$
\begin{equation}
    \label{cst:tvc} \left|\left|\begin{bmatrix}
        1 & 0 & 0 \\ 0 & 1 & 0
    \end{bmatrix}u_{B,R}(t) \right|\right|_2 \leq \tan(\theta)\ u_{B,R}(t)^T  \begin{bmatrix} 0 \\ 0 \\ 1\end{bmatrix}
\end{equation}
less than the maximum thrust,
\begin{equation}
    \label{cst:thrmax} ||u_{B,R}(t)||_2^2 \leq u_\text{max}
\end{equation}
and greater than the minimum thrust for the propulsive trajectory.
\begin{equation}
    \label{cst:thrmin}  u_\text{min} \leq ||u_{B,R}(t)||_2^2
\end{equation}

\noindent We impose a glideslope constraint of angle $\gamma$ to the propulsive component trajectory.
\begin{equation}
    \label{cst:gs} \left|\left|\begin{bmatrix}
        1 & 0 & 0 \\ 0 & 1 & 0
    \end{bmatrix}r_{I}(t)\right|\right|_2 \leq \tan(\gamma)\ r_I(t)^T  \begin{bmatrix} 0 \\ 0 \\ 1\end{bmatrix}
\end{equation}

\noindent We enforce a nonlinear net angle of attack with rolloff constraint:
\begin{equation}
    \label{cst:alphamax} \tanh \left(\frac{||v_{I}(t)||_2}{v_\text{small}} \right)\alpha(t) \leq \alpha_\text{max}
\end{equation}
to ensure lookup table validity. We use a hyperbolic rolloff at velocities below $v_\text{small}$ as the calculation of $\alpha$ has a singularity at $\|v\|_2 = 0$. As a result, numerical error near landing can therefore create very large yet practically meaningless values of $\alpha$.

\noindent Similarly, we enforce a maximum angular rate constraint
\begin{equation}
    \label{cst:omegamax} ||\omega_B(t)||_2 \leq \omega_\text{max}
\end{equation}
to limit centrifugal loads on the vehicle's structure.

\noindent Additionally, we impose maximum dynamic pressure and maximum dynamic pressure-angle of attack product limits:
\begin{align}
    \label{cst:qmax} \frac{1}{2}\rho(t)||v_I(t)||_2^2 &\leq q_\text{max} \\
    \label{cst:qamax} \frac{1}{2}\rho(t)\alpha(t)||v_I(t)||_2^2 &\leq \chi_\text{max}.
\end{align}

\section{Methods}

We aim to identify the reachable volume for the transition point between unpowered aerodynamic flight and powered descent guidance for a resuable launch vehicle. The~\ac{5DOF} dynamics and constraints we have described are highly nonlinear and the system has excessive dimensionality for traditional state-space exploration methods such as Hamilton-Jacobi methods while the imposed constraints make it unsuitable for set propagation or sampling-based methods.

The key observation is that we do not care about nonposition state at the ignition points so long as all dynamic constraints remain satisfied. We can thus avoid the curse of dimensionality if we can avoid examining vehicle attitude as part of our reachability analysis. We do so using a successive convexification-based trajectory optimization approach to ``fill in'' for the unconsidered states.

Unlike trajectory optimization based approaches for convex problems~\cite{eren2015constrained}, we cannot provide a guarantee that our computed sets over or underapproximates the reachable space. The reachable set for such a nonlinear problem may exhibit ``holes,'' concavities, consist of several disconnected sets, or have protuberances. Our algorithm cannot certify that the produced volume is either a super or subset of the true reachable space. What the approach does guarantee is that all of the explored points are actually reachable; we can always identify a feasible trajectory through a given previously identified point, but cannot always do so for interpolations thereof.

We first describe our method for solving the trajectory optimization subproblems, then examine our approach for using them to approximate the reachable volume.

\paragraph{Trajectory Optimization}

Our basic trajectory optimization subproblem is

\vspace{0.25cm}
\noindent\fbox{%
\parbox{0.98\linewidth}{%
\vspace{0.5em}
\centering{\text{\underline{Problem 1: Reference nonlinear trajectory optimization problem}}}
\begin{equation}
\tag{\textbf{P1}}
\begin{aligned}
      \min_{x,u} &\quad L(x, u) \\
    \text{ s.t. }
    &\quad \text{Eq.~\eqref{eqn:dyn:m},~\eqref{eqn:dyn:r},~\eqref{eqn:dyn:v},~\eqref{eqn:dyn:R},~\eqref{eqn:dyn:om},~\eqref{cst:aerocontrol}, ~\eqref{cst:ip}, ~\eqref{cst:fp}, ~\eqref{cst:mdry},} \\
    &\quad \text{~\eqref{cst:tvc}, \eqref{cst:thrmax}, ~\eqref{cst:thrmin}, ~\eqref{cst:gs}, ~\eqref{cst:alphamax}, ~\eqref{cst:omegamax}, ~\eqref{cst:qmax}, ~\eqref{cst:qamax} \thinspace.}
 \end{aligned} \label{pro:spbm_inf}
\end{equation}}}
\vspace{0.25cm}

Where $L(x, u)$ is a nonlinear objective function of the state and control trajectories.

We perform trajectory optimization by transcribing a nondimensionalized continuous-time model in ModelingToolkit.jl~\cite{ma2021modelingtoolkit} into a discrete transition system via multiple shooting which is then solved using the prox-linear method~\cite{elango2024} with successive convex optimization. Our implementation adopts the penalized trust region method using the exact $L_1$ penalty~\cite{bertsekas1975necessary} with the adaptive weighting scheme described by Mao et al~\cite{mao2018successive}. More details on the algorithm can be found in~\cite{malyuta2022convex}. 

Our multiple shooting method is the same as that used by Dukeman~\cite{dukeman2003enhancements} or Szmuk~\cite{szmuk2020successive}. The primary distinction is that we use two different time dilation parameters $\tau_a$ and $\tau_p$ (for aerodynamic and propulsive, respectively) to ``stretch'' nondimensionalized time $\tau$ into physical time $t$. Varying $\tau_a$ and $\tau_p$ then varies the overall flight time and the flight time of the segments as follows:
\[
t = \left\{ \begin{array}{cc}
    \tau_a \tau &: \tau < \frac{1}{2} \\
    \tau_p (\tau - \frac{1}{2}) + \frac{1}{2} \tau_a &: \tau \geq \frac{1}{2}
\end{array}\right.
\]
The total time of flight is then $\frac{1}{2} \tau_a + \frac{1}{2} \tau_p$. 

Aerodynamic controls $u_{W, a}(t)$ and propulsive controls $u_{B,R}(t)$ are transcribed using first-order-hold. Propulsive control is only applied when $\tau > \frac{1}{2}$; otherwise $u_{B,R}(\tau)=0$ where $\tau < \frac{1}{2}$. 

Constraints are either enforced nodally as part of the convex optimization problem \eqref{cst:ip},~\eqref{cst:fp},~\eqref{cst:mdry},~\eqref{cst:tvc},~\eqref{cst:thrmax},~\eqref{cst:gs},~\eqref{cst:omegamax} or through continuous time constraint satisfaction~\cite{elango2024} ~\eqref{cst:aerocontrol},~\eqref{cst:thrmin},~\eqref{cst:alphamax},~\eqref{cst:omegamax},~\eqref{cst:qamax},~\eqref{cst:qmax}. We transcribe Eq.~\eqref{cst:omegamax} both as a second order cone constraint on each node to improve convergence of the low frequency dynamics and with~\ac{CTCS} to reduce inter-sample constraint violation caused by the high frequency dynamics. Additionally, while Eq.~\ref{cst:aerocontrol} is a control constraint, it strongly depends on rapidly changing vehicle states so we therefore transcribe it using~\ac{CTCS}. Unlike Elago et al~\cite{elango2024}, we convert each~\ac{CTCS} constraint as an independent state variable in our augmented dynamics to improve the gradients of the sensitivity problems.

Most of the computation for the prox-linear method is incurred in sensitivty analysis of the integrated multiple shooting subproblems. We compute the Jacobians for each of the subproblems in parallel using SciMLSensitivity~\cite{rackauckas2020universal} optimized for sparsity using SparseConnecitvityTracer~\cite{hill_2024_13961066} which are then solved on a single thread using the Clarabel optimizer~\cite{Clarabel_2024} via JuMP's disciplined convex programming interface~\cite{Lubin2023}. 

\paragraph{Reachability Analysis}

We use a naive version of the defect hull algorithm proposed by Chan~\cite{chan2022optimization} to perform reachability analysis. We first compute a reference trajectory using a min-fuel objective begetting an ignition point $n_{I, ip}$. The reachability polytope $P_i$ is then initialized from this singleton.

\begin{figure}
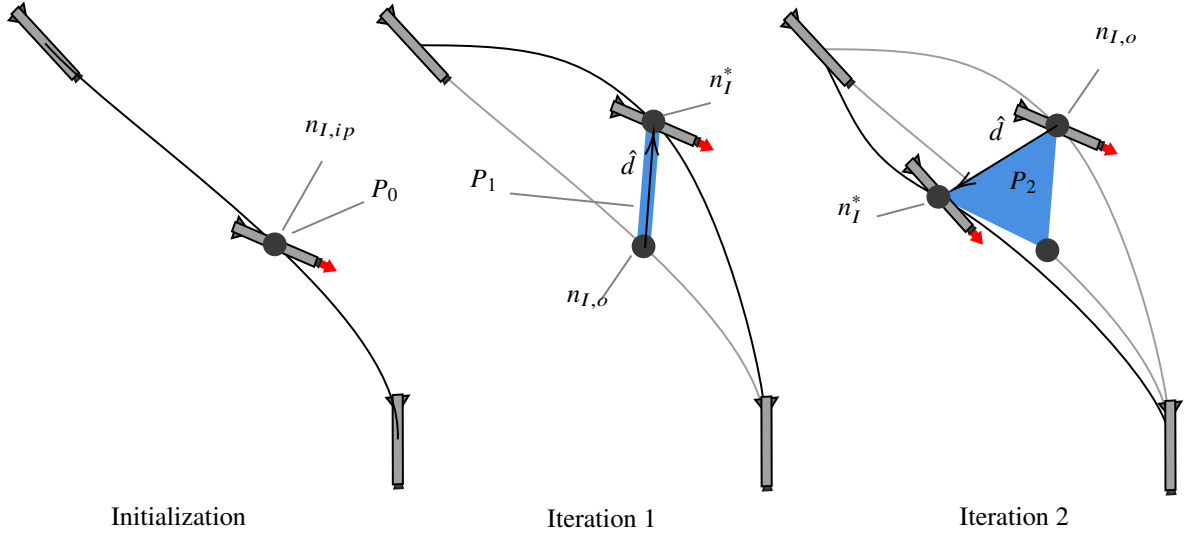

\centering
\include{algo}
\caption{Schematic of reachability algorithm iteration}
\label{fig:reachability-algo}
\end{figure}

At each iteration the algorithm, depicted in Figure~\ref{fig:reachability-algo} picks a point $n_{I,o}$ on the surface of $P_i$ and selects a random direction $\hat{d}$ to serve as an expansion vector. It then solves the defect hull problem:

\vspace{0.25cm}
\noindent\fbox{%
\parbox{0.98\linewidth}{%
\vspace{0.5em}
\centering{\text{\underline{Problem 2: Defect hull trajectory optimization problem}}}
\begin{equation}
\tag{\textbf{P2}}
\begin{aligned}
      \min_{x,u} &\quad -\mu \\
    \text{ s.t. }
    &\quad n_{I} = r_I(0.5) = \mu \hat{d} + n_{I,o} \\
    &\quad \text{Eq.~\eqref{eqn:dyn:m},~\eqref{eqn:dyn:r},~\eqref{eqn:dyn:v},~\eqref{eqn:dyn:R},~\eqref{eqn:dyn:om},~\eqref{cst:aerocontrol}, ~\eqref{cst:ip}, ~\eqref{cst:fp}, ~\eqref{cst:mdry},} \\
    &\quad \text{~\eqref{cst:tvc}, \eqref{cst:thrmax}, ~\eqref{cst:thrmin}, ~\eqref{cst:gs}, ~\eqref{cst:alphamax}, ~\eqref{cst:omegamax}, ~\eqref{cst:qmax}, ~\eqref{cst:qamax} \thinspace.}
 \end{aligned} \label{pro:defecthull}
\end{equation}}}
\vspace{0.25cm}

The ensuing solution $n_I^*$ is then the ignition point farthest from $n_{I,o}$ along $\hat{d}$ that retains feasibility with respect to our state and control constraints. The reachability polytope is then expanded to $P_{i+1} = \texttt{convexhull}(P_{i}, n_I^*)$ and the process is repeated.

\section{Numerical Example}

\begin{figure}
\begin{floatrow}
\ffigbox{%
\includegraphics[scale=0.2]{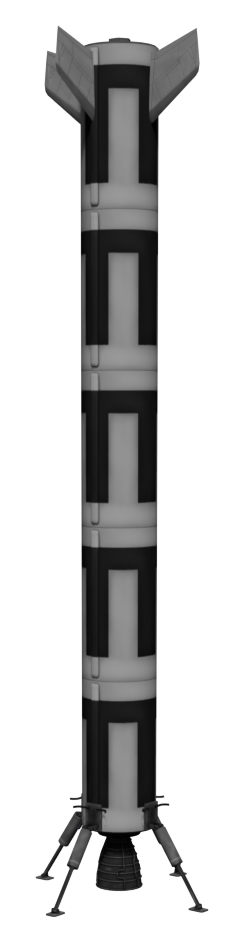}
}{
\caption{Example reusable launch vehicle from Kerbal Space Program.}
\label{fig:kerbal}
}
\capbtabbox{%
    \begin{tabular}{|c|c|}
    \hline
        $m_\text{dry}$ & \qty{10,088}{\kilogram} \\
        $m_\text{wet}$ & \qty{19,516}{\kilogram} \\ 
        $J_{B,\text{dry}}$ & $\text{diag}(4.4, 4.4, 0.040) \times 10^5\ \unit{\kilogram\per\meter\squared}$\\
        $J_{B,\text{wet}}$ & $\text{diag}(5.6, 5.6, 0.053) \times 10^5\ \unit{\kilogram\per\meter\squared}$\\
        $I_\text{sp}$ & \qty{300}{\second} \\ 
        $u_\text{max}$ & \qty{936}{\kilo\newton} \\
        $r_{B,\text{engine}}$ & $[0;0;-4.1]\unit{\meter}$ \\
        $r_{B,\text{fin}}$ & $[0;0;9.2]\unit{\meter}$ \\
        $\theta_\text{max}$ & $\SI{10.5}{\degree}$ \\
        $\omega_\text{max}$ & $\SI{15}{\degree}$ \\
        $C_\text{damp}$ & 10 \\
        \hline
    \end{tabular}
}{%
    \caption{Vehicle parameters}
    \label{tab:vehparams}
}
\end{floatrow}
\end{figure}

Our example vehicle is a simple vertical takeoff vertical landing reusable launch vehicle in the medium-fidelity~\ac{6DOF} simulator~\ac{KSP} using the Ferram Aerospace Research aerodynamics extension. The detailed parameters are in Table~\ref{tab:vehparams}, where $\text{diag}(v)$ refers to the matrix with $v$ on the diagonal.

Our reference scenario depicts a return to launch site trajectory for the first stage on the fictional planet Kerbin, which imposes several challenges compared to a similar scenario on Earth:
\begin{itemize}
    \item Kerbin is small. Kerbin's radius ($\qty{600}{\kilo\meter}$) is similar to Charon's. As a result, the tangent plane approximation breaks down quickly, forcing us to use a spherical altitude model.
    \item Kerbin rotates very quickly. Kerbin completes a rotation about its axis in only 6 hours, causing rotating reference frame effects to become quickly noticeable.
    \item Kerbin is extremely dense. In spite of its small size, Kerbin's surface gravity is approximately the same as Earth's. The gravitational field therefore falls off very quickly, thus causing our use of a spherical gravity model.
    \item Kerbin has 20\% shorter atmosphere than Earth's, reducing the opportunity for aerodynamic maneuvering.
\end{itemize}
All four effects would also occur in a similar trajectory on Earth but with smaller magnitudes. The most important effects of Kerbin's odd dimensions are the latter two; Kerbin's rapid gravitational falloff causes errors of approximately $\qty{60}{\meter\per\second}$ over trajectories of our duration and its shrunk atmosphere causes peak aerodynamic authority to happen very close to engine ignition for our trajectories.

Our vehicle was constructed using stock parts in~\ac{KSP}, consisting of five fuel tanks, landing legs, and fin actuators. Our return scenario depicts only the bottom-most fuel tank being full, with the others empty.~\ac{KSP} models fuel drain in the stage as occurring from all tanks in the stage at once therein causing the moment of inertia and center of mass to move rapidly towards the engine as fuel is depleted. By pre-draining the upper tanks, we approximate the real effect of ullage on the center of mass and moment of inertia.

We extracted aerodynamic data from the game using a custom version of the Kerbal Remote Procedure Call extension that allowed us to query the aerodynamic forces and torques on the vehicle at different velocities, altitudes, and control configurations. To extract the data needed to build the aerodynamic lookup tables, we performed sweeps over each angle of attack axis up to 25 degrees, each plane of aerodynamic control actuation up to 45 degrees, and in mach number up to mach 3.0. The data at each angle of attack was referenced to the control condition of minimum aerodynamic force (corresponding to when each actuator was pointed into the wind vector) and the body forces and moments were then extracted. The drag polar for the fin actuators was then developed from a least squares fit to the lower surface of a convex hull at this condition. 

All code, including our dynamics, convex subproblem solver, reachability analysis implementation, and plot generation utilities, as well as the modified Kerbal Remote Procedure Call extension and our post-processing scripts are available in the \href{Github}{https://github.com/BenChung/DualPhaseDescent.jl} repository. Additionally, we include raw and processed aerodynamic data for our vehicle alongside the~\ac{KSP} craft file in the repository. The version of the code and supporting material as submitted is on Git branch \texttt{SciTech2025}. 

The parameters for our reference return problem are given in Table~\ref{tab:trajparams}. The problem was chosen to be comparable to the post-retropropulsion trajectory from Falcon 9 missions~\cite{doi:10.2514/1.A34486}. 

\begin{figure}
\begin{floatrow}
\capbtabbox{%
    \begin{tabular}{|c|c|}
    \hline
        $S_{\alpha.\text{max}}$ & $1$ \\
        $S_{\text{thrust},\text{min}}$ & $20$ \\
        $S_{\text{fin},\text{bound}}$ & $10$ \\
        $S_{||\omega||_2,\text{max}}$ & $0.1$ \\
        $S_{q,\text{max}}$ & $5\times 10^{-4}$ \\
        $S_{q\alpha,\text{max}}$ & $1\times 10^{-6}$ \\
        \hline
    \end{tabular}
}{
\caption{CTCS constraint scaling.}
\label{tab:cweights}
}
\capbtabbox{%
    \centering
    \begin{tabular}{|c|c|}
    \hline
        $r_I$ & $[0.5; 2.5; 15]\unit{\kilo\meter}$ \\
        $v_I$ & $[0; -150; -350]$m/s \\
        $R_{B/I, i}$ & $[-0.98; 0]$ \\
        $\omega_{B, i}$ & $[0; 0]\unit{\radian\per\second}$ \\
        $r_f$ & $[0; 0;0]\unit{\meter}$ \\
        $v_f$ & $[ 0; 0; 0]\unit{\meter\per\second}$ \\
        $R_{B/I, f}$ & $[0; 0]\unit{\degree}$ \\
        $\omega_{B, f}$ & $ [ 0; 0]\unit{\radian\per\second}$ \\
        $\gamma$ & $\SI{60}{\degree}$ \\
        $v_\text{small}$ & $\SI{100}{m/s}$ \\
        $\rho_\text{max}$ & $\SI{8e4}{\pascal}$ \\
        $\chi_\text{max}$ & $\SI{1e6}{\pascal\degree}$ \\
        \hline
    \end{tabular}
}{%
    \caption{Trajectory parameters}
    \label{tab:trajparams}
}
\end{floatrow}  
\end{figure}

\paragraph{Initialization Problem}

\begin{figure}
\begin{floatrow}
\capbtabbox{%
    \hspace{3em}
    \centering
    \begin{tabular}{|c|c|}
    \hline
        $N$ & $41$ \\
        $\beta$ & $2.0$ \\
        $\alpha$ & $2.0$\\
        $\rho_0$ & $0.0$ \\
        $\rho_1$ & $0.25$ \\
        $\rho_2$ & $0.7$ \\
        $r_\text{init}$ & $8.0$ \\
        $w_m$ & $1000$ \\
        $w_n$ & $50$ \\
        $w_l$ & $100$ \\
        \hline
    \end{tabular}
    \hspace{3em}
}{%
    \caption{Optimizer parameters}
    \label{tab:optparams}
}
\ffigbox{%
\includegraphics[scale=0.5]{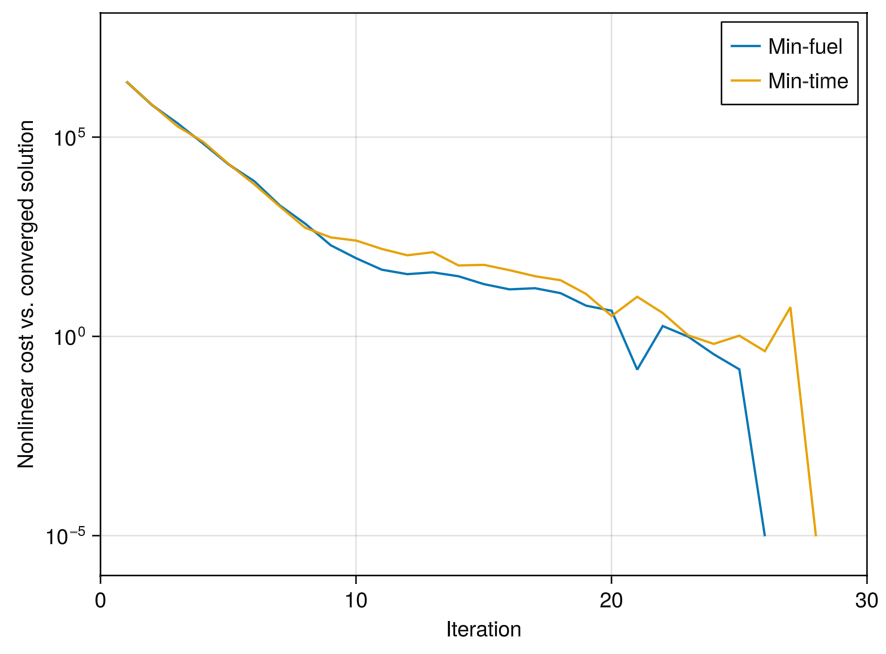}
}{
\caption{Convergence histories for the min-fuel and min-time problems.}
\label{fig:convergence}
}
\end{floatrow}
\end{figure}

To initialize the reachability analysis we need a reference trajectory from which to start exploring the feasible ignition points. For illustration, we will show both a min-fuel and min-time version of this problem. SCP is initialized from a zero-control trajectory propagated forward from the initial condition; for this section, we set our tolerance at 1e-5. Table~\ref{tab:cweights} shows the CTCS penalty weights we used for each constraint applied in continuous time. Optimizer parameters are shown in table~\ref{tab:optparams}; starting from $r_\text{init}$ we expand and contract the trust region weight by $\alpha$ and $\beta$ respectively when the relative improvement is better than $\rho_2$ or smaller than $\rho_1$, and reject the iterate if it is less than $\rho_0$. For convergence, we apply the penalties $w_m$ to the linearization error of the dynamics, $w_n$ to the linearization error of the terminal constraint, and $w_l$ on the CTCS equality state violation at each iteration.

Convergence histories for the initialization problems are shown in Figure~\ref{fig:convergence}.

\begin{figure}
    \centering
    \includegraphics[width=1.3\linewidth,angle=-90]{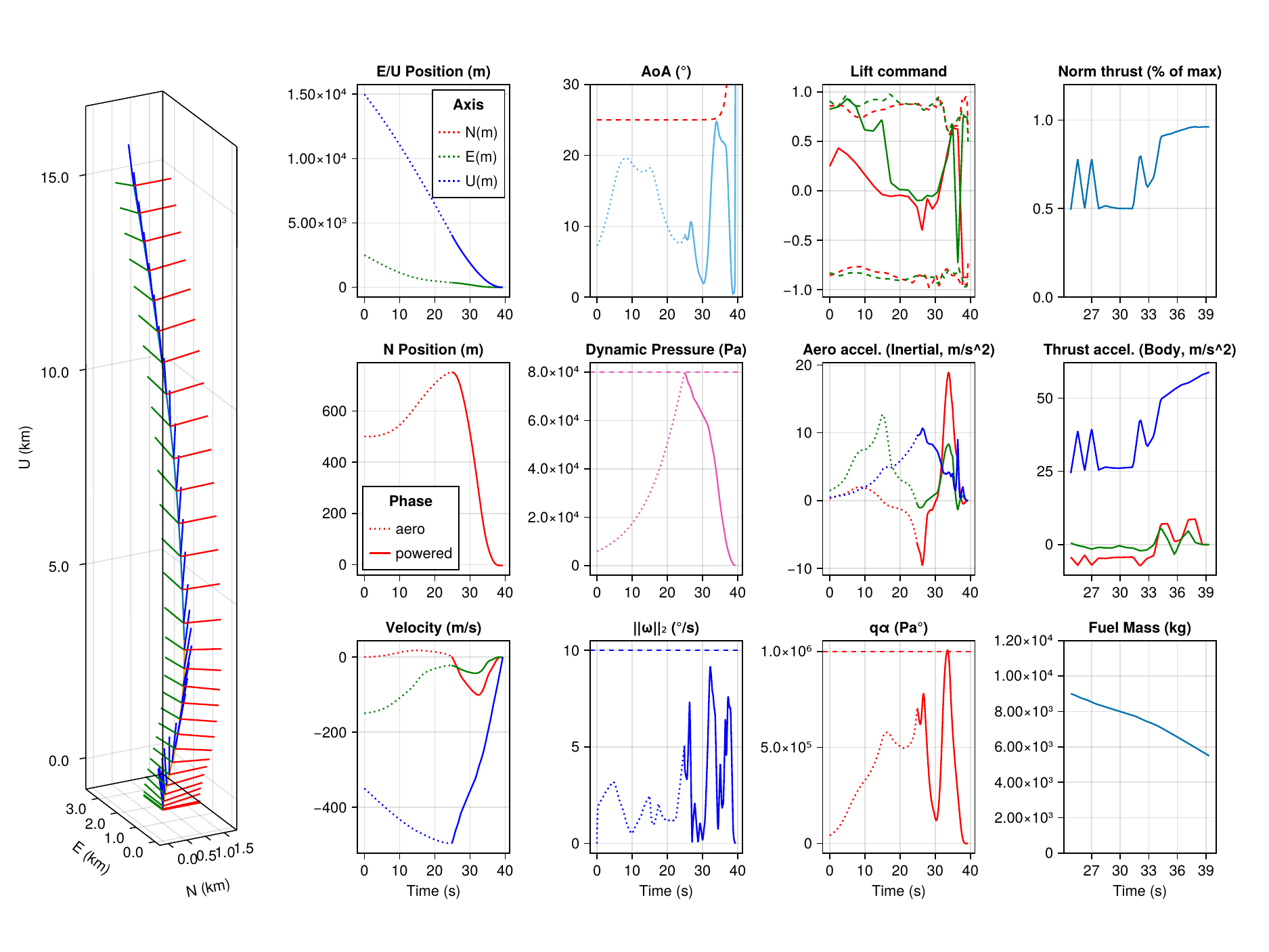}
    \caption{Initialization min-time trajectory}
    \label{fig:min-fuel-init}
\end{figure}

We show a min-fuel solution in Figure~\ref{fig:min-fuel-init}. Two features stand out compared to vacuum or lift-free formulations such as~\cite{szmuk2020successive}:
\begin{itemize}
    \item The rocket maneuvers (``swerves'') in the latter phases of the aerodynamic trajectory, reducing the energy going into the propulsive phase.
    \item The powered descent is done at nonzero angle of attack until the last phases to use lift to non-propulsively cancel velocity.
\end{itemize}
These features are unique to coupled aerodynamic-propulsive optimization; without combining the effects the optimizer would be unable to trade off position for reduced velocity and therefore reduced fuel consumption.

We would furthermore like to highlight the utility of the~\ac{CTCS} transcription for this problem. The damped attitude dynamics of the rocket are still much faster than the control or state discretization and are very wont to create inter-sample constraint violation. By using~\ac{CTCS}, we can minimize this inter-sample constraint violation without needing to refine our optimization time grid. In particular, we highlight the instantaneous constraint activation shown in the dynamic pressure and q$\alpha$ histories. 

The no-stall, travel-limited aerodynamic actuator restriction imposed by Eq.~\eqref{cst:aerocontrol} are also tracked well with the~\ac{CTCS} method, with the history shown in the plot of lift commands over the trajectory, where the trajectory tracks a rapidly changing limit with one or both of its aerodynamic controls.

\paragraph{Reachability}

\begin{figure}
\begin{floatrow}
\ffigbox{%
\includegraphics[scale=0.7]{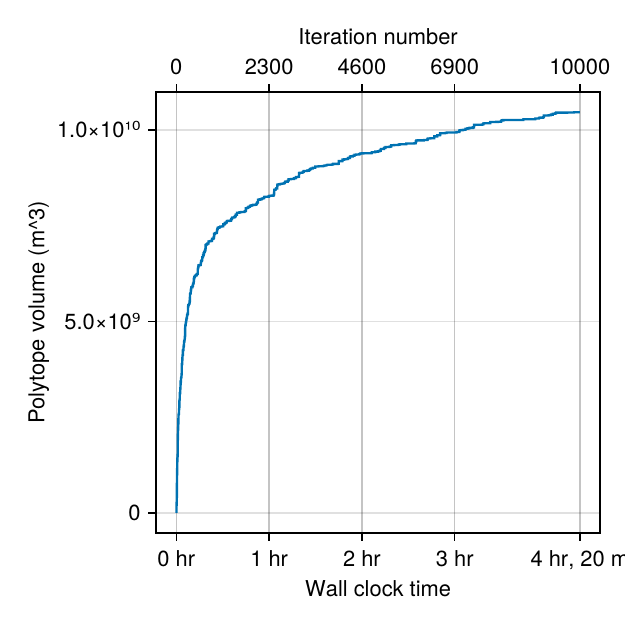}
}{
\caption{Convergence history and wall clock time for the reachability problem.}
\label{fig:convergence}
}
\ffigbox{%
\includegraphics[scale=0.7]{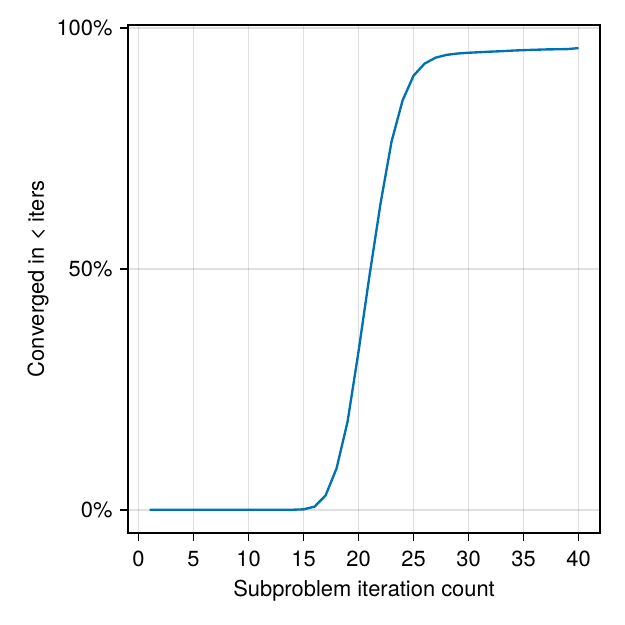}
}{
\caption{Subproblem convergence history.}
\label{fig:spbmconvergence}
}
\end{floatrow}
\end{figure}

\begin{figure}
\begin{floatrow}
\ffigbox{%
\includegraphics[scale=0.7]{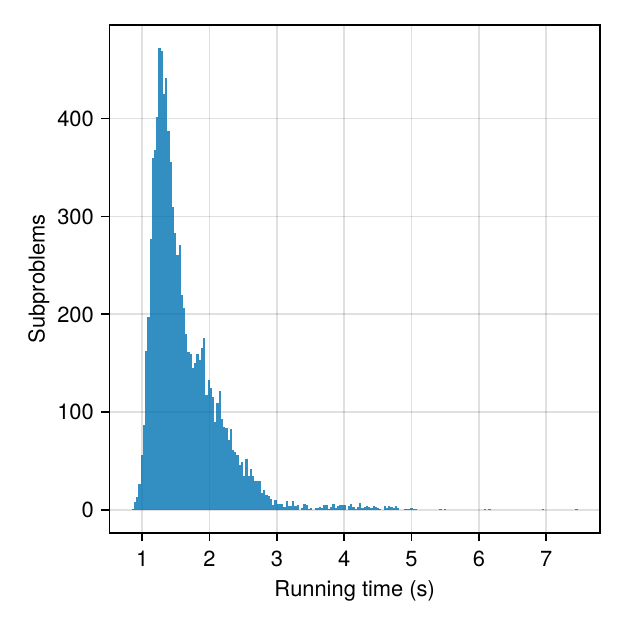}
}{
\caption{Subproblem running time distribution.}
\label{fig:runtimes}
}
\ffigbox{%
\includegraphics[scale=0.7]{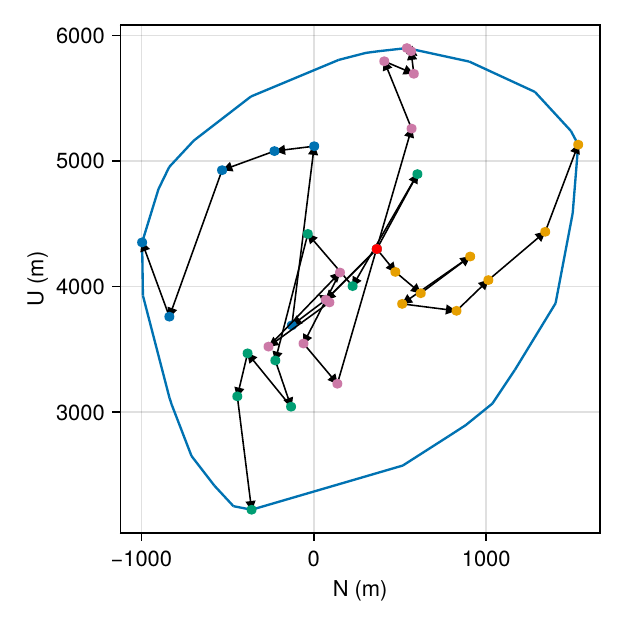}
}{
\caption{Expansion trajectories for extremal points on the polytope. Intermediate points are on the surface of the polytope at the time they were chosen, but may not have been the extremal points on the associated axis.}
\label{fig:expandtraj}
}
\end{floatrow}
\end{figure}

We then initialize the reachability analysis from the min-fuel ignition point and expand the hull for 10,000 iterations. Figure~\ref{fig:convergence} shows the convergence rate as a function of polytope volume through the reachability problem by iteration and by wall clock time. Figure~\ref{fig:spbmconvergence} shows the cumulative convergence of each subproblem by SCP iteration. The curve does not reach $\SI{100}{\%}$ since we reject iterations whose solutions were not dynamically feasible; we rejected $\SI{4.3}{\%}$ of generated solutions for dynamic infeasibility and $\SI{0.02}{\%}$ due to numerical instabilities encountered during the solve. 
Total elapsed time was 4 hours, 20 minutes running on a AMD Ryzen AI 9 HX 370 with 32 threads, for an average time per defect hull trajectory optimization subproblem of 1.63 seconds; the histogram of running time is shown in Figure~\ref{fig:runtimes}.

\begin{figure}
    \centering
\includegraphics[width=\linewidth]{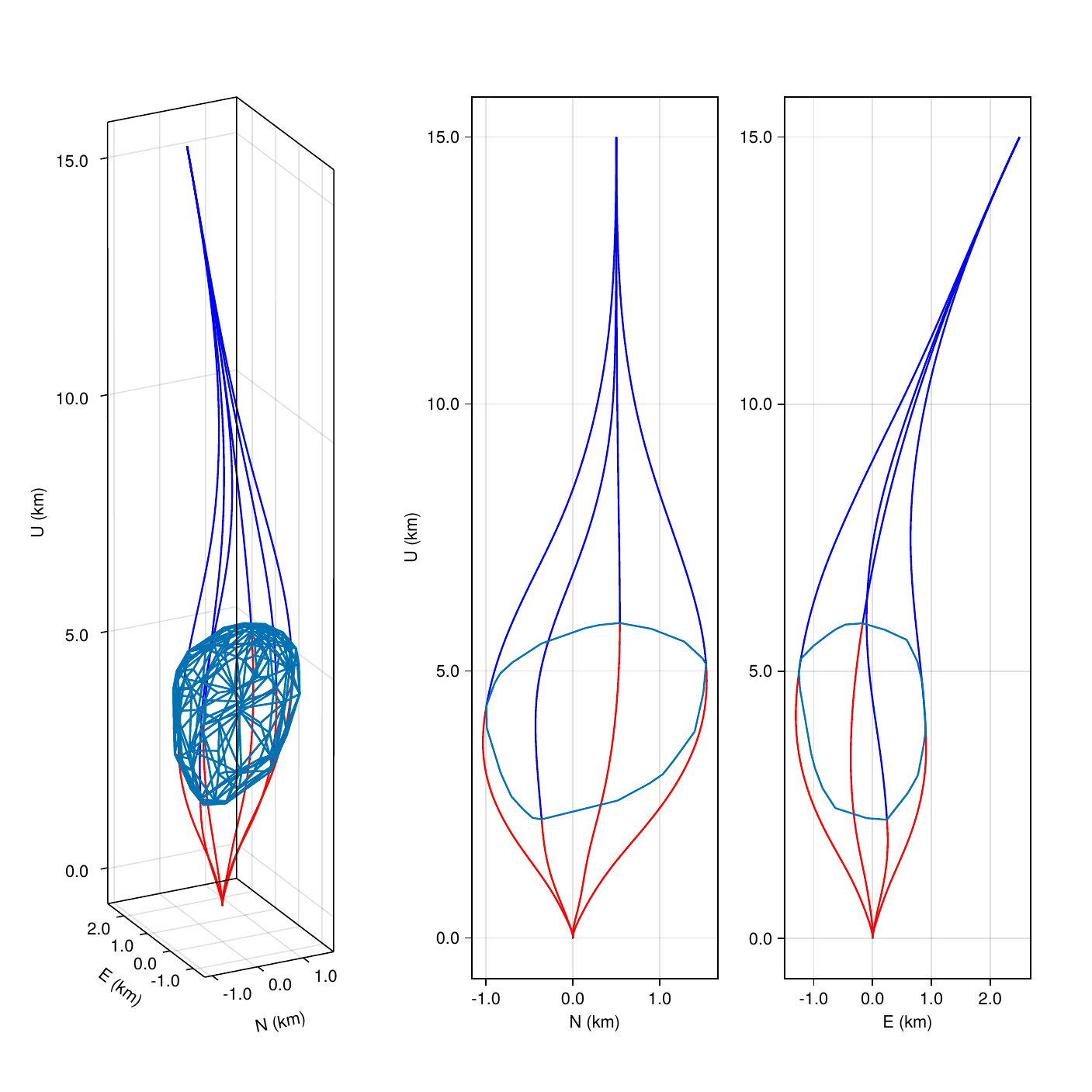}
    \caption{Reachable ignition polytope with extremal trajectories}
    \label{fig:reachable}
\end{figure}

The reachable polytope computed for our example problem is shown in Figure~\ref{fig:reachable}. We visualize only the extremal (farthest in each cardinal direction) trajectories in each plane for the orthogonal views. We visualize the route of sampled points taken to construct the extremal points inn the north-up plane in Figure~\ref{fig:expandtraj}, starting from the ignition point of the initialization trajectory in red and expanding outwards to the surface of the final polytope.

The progress of the polytope expanding shown in Figure~\ref{fig:expandtraj} illustrates the nonconvexity of the subproblem. The subproblems make small steps inside of the reachable volume instead of immediately expanding to the reachable frontier. This slow progress is caused by the individual defect hull tasks falling into local minima based on their initialization and solution trajectory. The reachability algorithm is then able to expand further by re-initializing from a previously identified extreme point and escape the local minima only to find another one.

\section{Conclusion}
We described an approach for approximating the projection of a high-dimensional constrained reachability problem into a low dimensional space via nonlinear optimization. Our problem is not tractable for traditional reachability methods: not only does it have 11 state dimensions, it also has state and control constraints that set propagation based methods cannot easily accommodate. We avoid the curse of dimensionality by focusing only on the relevant low-dimensional subset of the greater reachable volume, thereby dramatically reducing the dimensionality of the space we are examining, then explore said volume using nonlinear optimization.

We show that the algorithm works to approximate the feasible ignition points for a 5 degree of freedom model of an aerodynamically controlled reusable launch vehicle's axisymmetric first stage while subject to many practical constraints. We see substantial opportunity for future work, as our approach described here was rather naive. While our algorithm was able to approximate the reachable volume, it took several hours to do so due to implementation limitations and sampling inefficiency. We think that several fruitful directions may consist of:
\begin{itemize}
    \item Customized sparse implementation. Around 30\% of our wall clock time was wasted due to convex parsing overheads; in particular, constructing the sparse constraint matrices and copying them into the solver's working memory. Using pre-allocated and customized sparse data structures and a solver optimized for such would provide an integer speedup. For example, Kamath et al.~\cite{kamath2023real} was able to calculate solutions to a similar nonconvex trajectory optimization problem in $\SI{10}{s}$ of ms, though with a considerably smaller state dimension.
    \item Improved SCP algorithm. New work, such as Auto-SCvx~\cite{mceowen2024arxiv, mceowen2025}, is likely to substantially reduce the iteration count for subproblem convergence; their problem, which has similar constraints to ours, solved in less than 7 iterations on average.
    \item GPU implementation. As~\cite{chan2022optimization} suggests, the defect hull algorithm is intrinsically very parallelizable, as the selection of initial points and expansion directions is stochastic and largely independent. Furthermore, the multiple shooting sensitivity analysis is an embarrassingly parallel process as each of the constituent trajectories can be simulated independently. Chari et al.~\cite{doi:10.2514/6.2024-1762} used a pure-GPU implementation of SCP and achieved up to 5x uplift compared to a CPU implementation. CPU-GPU pipelining of convex optimization and sensitivity analysis may yield further performance uplift.
    \item Our version of Chan's defect hull algorithm is very naive in its sampling strategy as it does not use the dual-derived tangent information that Chan's proposed algorithm does. We believe that our slow convergence rate in terms of number of iterations is substantially due to this limitation. Improving reachability sampling efficiency would then reduce the number of trajectories needed to achieve sufficient polytope volume convergence.
\end{itemize}
We believe that nonlinear trajectory optimization based reachability analysis can allow the approximation of projections of highly nonlinear reachability problems. In this paper, we demonstrated that the approach can work on a practical multiphase trajectory for a aerodynamically and propulsively controlled reusable launch vehicle by approximating the spatial volume in which its ignition point can be placed. Furthermore, we believe that there are opportunities for substantial further algorithmic and implementation optimizations above our algorithm that can yield overall computation times in the minutes or seconds, instead of hours.

Our approach may also be generalized onto other classes of problem that are tractable for trajectory optimization. One concept is to extremize a metric like conditional value-at-risk, which has already been explored extensively using SCP~\cite{kazu2024,kazu2024dissertation} which would then allow for reachability analysis over best and worst case outcomes for a stochastic system. Another would be to extend our two-phase analysis into a multiphase one derived from temporal and logical specifications~\cite{dgmsr} which would allow automatic identification of the state spaces in which transitions could occur.
 
\bibliography{acl,sample}

\begin{thebibliography}{29}
\newcommand{\enquote}[1]{``#1''}
\providecommand{\natexlab}[1]{#1}
\providecommand{\url}[1]{\texttt{#1}}
\providecommand{\urlprefix}{URL }
\expandafter\ifx\csname urlstyle\endcsname\relax
  \providecommand{\doi}[1]{\discretionary{}{}{}https://doi.org/#1}\else
  \providecommand{\doi}[1]{\discretionary{}{}{}\urlstyle{rm}\url{https://doi.org/#1}}\fi

\bibitem[{Blackmore(2016)}]{blackmore2016autonomous}
Blackmore, L., \enquote{Autonomous precision landing of space rockets,} \emph{The Bridge on Frontiers of Engineering}, Vol.~4, No.~46, 2016, pp. 15--20.

\bibitem[{Szmuk et~al.(2020)Szmuk, Reynolds, and A{\c{c}}{\i}kme{\c{s}}e}]{szmuk2020successive}
Szmuk, M., Reynolds, T.~P., and A{\c{c}}{\i}kme{\c{s}}e, B., \enquote{Successive convexification for real-time six-degree-of-freedom powered descent guidance with state-triggered constraints,} \emph{{AIAA Journal of Guidance, Control, and Dynamics}}, Vol.~43, No.~8, 2020, pp. 1399--1413.
\newblock \doi{10.2514/1.G004549}.

\bibitem[{Sagliano et~al.(2021)Sagliano, Heidecker, Hernández, Farì, Schlotterer, Woicke, Seelbinder, and Dumont}]{doi:10.2514/6.2021-0862}
Sagliano, M., Heidecker, A., Hernández, J.~M., Farì, S., Schlotterer, M., Woicke, S., Seelbinder, D., and Dumont, E., \enquote{Onboard Guidance for Reusable Rockets: Aerodynamic Descent and Powered Landing,} \emph{{AIAA Scitech Forum}}, 2021.
\newblock \doi{10.2514/6.2021-0862}.

\bibitem[{Bansal et~al.(2017)Bansal, Chen, Herbert, and Tomlin}]{bansal2017hamilton}
Bansal, S., Chen, M., Herbert, S., and Tomlin, C.~J., \enquote{Hamilton-jacobi reachability: A brief overview and recent advances,} \emph{{Proc.\ IEEE Conf.\ on Decision and Control}}, 2017.
\newblock \doi{10.1109/CDC.2017.8263977}.

\bibitem[{Althoff et~al.(2021)Althoff, Frehse, and Girard}]{althoff2021set}
Althoff, M., Frehse, G., and Girard, A., \enquote{Set propagation techniques for reachability analysis,} \emph{{Annual Review of Control, Robotics, and Autonomous Systems}}, Vol.~4, 2021, pp. 369--395.
\newblock \doi{10.1146/annurev-control-071420-081941}.

\bibitem[{Lew and Pavone(2020)}]{lew2021sampling}
Lew, T., and Pavone, M., \enquote{Sampling-based reachability analysis: A random set theory approach with adversarial sampling,} \emph{{Conference on Robot Learning}}, 2020.
\newblock \doi{10.48550/arXiv.2008.10180}.

\bibitem[{Dueri et~al.(2014)Dueri, A{\c{c}}{\i}kme{\c{s}}e, Baldwin, and Erwin}]{dueri2014finite}
Dueri, D., A{\c{c}}{\i}kme{\c{s}}e, B., Baldwin, M., and Erwin, R.~S., \enquote{Finite-horizon controllability and reachability for deterministic and stochastic linear control systems with convex constraints,} \emph{{American Control Conference}}, 2014.
\newblock \doi{10.1109/ACC.2014.6859302}.

\bibitem[{Chan(2022)}]{chan2022optimization}
Chan, K.~W., \enquote{Optimization-Based Reachability Analysis for Landing Scenarios,} Ph.D. thesis, Universit{\"a}t Bremen, 2022.
\newblock \doi{10.26092/elib/1922}.

\bibitem[{ISO 2533:1975()}]{ISO2533}
ISO 2533:1975, \enquote{{Standard Atmosphere},} Standard, International Organization for Standardization, Geneva, CH, Mar. 1975.

\bibitem[{Dukeman and Calise(2003)}]{dukeman2003enhancements}
Dukeman, G., and Calise, A., \enquote{Enhancements to an atmospheric ascent guidance algorithm,} \emph{{AIAA Guidance, Navigation, and Control Conference and Exhibit}}, 2003.
\newblock \doi{10.2514/6.2003-5638}.

\bibitem[{Möller and Hughes(1999)}]{doi:10.1080/10867651.1999.10487509}
Möller, T., and Hughes, J.~F., \enquote{Efficiently Building a Matrix to Rotate One Vector to Another,} \emph{Journal of Graphics Tools}, Vol.~4, No.~4, 1999, pp. 1--4.
\newblock \doi{10.1080/10867651.1999.10487509}.

\bibitem[{Eren et~al.(2015)Eren, Dueri, and A{\c{c}}{\i}kme{\c{s}}e}]{eren2015constrained}
Eren, U., Dueri, D., and A{\c{c}}{\i}kme{\c{s}}e, B., \enquote{Constrained reachability and controllability sets for planetary precision landing via convex optimization,} \emph{{AIAA Journal of Guidance, Control, and Dynamics}}, Vol.~38, No.~11, 2015, pp. 2067--2083.
\newblock \doi{10.2514/1.G000882}.

\bibitem[{Ma et~al.(2021)Ma, Gowda, Anantharaman, Laughman, Shah, and Rackauckas}]{ma2021modelingtoolkit}
Ma, Y., Gowda, S., Anantharaman, R., Laughman, C., Shah, V., and Rackauckas, C., \enquote{ModelingToolkit: A Composable Graph Transformation System For Equation-Based Modeling,} , 2021.
\newblock {Available at }\url{https://arxiv.org/abs/2103.05244}.

\bibitem[{Elango et~al.(2024)Elango, Luo, Kamath, Uzun, Kim, and A{\c{c}}ikmese}]{elango2024}
Elango, P., Luo, D., Kamath, A.~G., Uzun, S., Kim, T., and A{\c{c}}ikmese, B., \enquote{Successive Convexification for Trajectory Optimization with Continuous-Time Constraint Satisfaction,} , 2024.
\newblock {Available at }\url{https://arxiv.org/abs/2404.16826}.

\bibitem[{Bertsekas(1975)}]{bertsekas1975necessary}
Bertsekas, D.~P., \enquote{Necessary and sufficient conditions for a penalty method to be exact,} \emph{{Mathematical Programming}}, Vol.~9, No.~1, 1975, pp. 87--99.
\newblock \doi{10.1007/BF01681332}.

\bibitem[{Mao et~al.(2018)Mao, Szmuk, Xu, and A{\c{c}}ikmese}]{mao2018successive}
Mao, Y., Szmuk, M., Xu, X., and A{\c{c}}ikmese, B., \enquote{Successive convexification: A superlinearly convergent algorithm for non-convex optimal control problems,} , 2018.
\newblock {Available at }\url{https://arxiv.org/abs/1804.06539}.

\bibitem[{Malyuta et~al.(2022)Malyuta, Reynolds, Szmuk, Lew, Bonalli, Pavone, and A{\c{c}}{\i}kme{\c{s}}e}]{malyuta2022convex}
Malyuta, D., Reynolds, T.~P., Szmuk, M., Lew, T., Bonalli, R., Pavone, M., and A{\c{c}}{\i}kme{\c{s}}e, B., \enquote{Convex optimization for trajectory generation: A tutorial on generating dynamically feasible trajectories reliably and efficiently,} \emph{{IEEE Control Systems Magazine}}, Vol.~42, No.~5, 2022, pp. 40--113.
\newblock \doi{10.1109/MCS.2022.3187542}.

\bibitem[{Rackauckas et~al.(2020)Rackauckas, Ma, Martensen, Warner, Zubov, Supekar, Skinner, and Ramadhan}]{rackauckas2020universal}
Rackauckas, C., Ma, Y., Martensen, J., Warner, C., Zubov, K., Supekar, R., Skinner, D., and Ramadhan, A., \enquote{Universal differential equations for scientific machine learning,} , 2020.
\newblock {Available at }\url{https://arxiv.org/abs/2001.04385}.

\bibitem[{Hill and Dalle(2024)}]{hill_2024_13961066}
Hill, A., and Dalle, G., \enquote{SparseConnectivityTracer.jl,} , Oct. 2024.
\newblock \doi{10.5281/zenodo.13961066}, \urlprefix\url{https://doi.org/10.5281/zenodo.13961066}.

\bibitem[{Goulart and Chen(2024)}]{Clarabel_2024}
Goulart, P.~J., and Chen, Y., \enquote{Clarabel: An interior-point solver for conic programs with quadratic objectives,} , 2024.
\newblock {Available at }\url{https://arxiv.org/abs/2405.12762}.

\bibitem[{Lubin et~al.(2023)Lubin, Dowson, {Dias Garcia}, Huchette, Legat, and Vielma}]{Lubin2023}
Lubin, M., Dowson, O., {Dias Garcia}, J., Huchette, J., Legat, B., and Vielma, J.~P., \enquote{{JuMP} 1.0: {R}ecent improvements to a modeling language for mathematical optimization,} \emph{{Mathematical Programming Computation}}, Vol.~15, No.~3, 2023.
\newblock \doi{10.1007/s12532-023-00239-3}.

\bibitem[{Ecker et~al.(2020)Ecker, Karl, Dumont, Stappert, and Krause}]{doi:10.2514/1.A34486}
Ecker, T., Karl, S., Dumont, E., Stappert, S., and Krause, D., \enquote{Numerical Study on the Thermal Loads During a Supersonic Rocket Retropropulsion Maneuver,} \emph{{AIAA Journal of Spacecraft and Rockets}}, Vol.~57, No.~1, 2020, pp. 131--146.
\newblock \doi{10.2514/1.A34486}.

\bibitem[{Kamath et~al.(2023)Kamath, Elango, Yu, Mceowen, Chari, Carson~III, and A{\c{c}}{\i}kme{\c{s}}e}]{kamath2023real}
Kamath, A.~G., Elango, P., Yu, Y., Mceowen, S., Chari, G.~M., Carson~III, J.~M., and A{\c{c}}{\i}kme{\c{s}}e, B., \enquote{Real-Time Sequential Conic Optimization for Multi-Phase Rocket Landing Guidance,} \emph{{IFAC World Congress}}, 2023.
\newblock \doi{https://doi.org/10.1016/j.ifacol.2023.10.1444}.

\bibitem[{Mceowen et~al.(2024)Mceowen, Calderone, Tiwary, Zhou, Kim, Elango, and A{\c{c}}ikmese}]{mceowen2024arxiv}
Mceowen, S., Calderone, D.~J., Tiwary, A., Zhou, J. S.~K., Kim, T., Elango, P., and A{\c{c}}ikmese, B., \enquote{Auto-tuned Primal-dual Successive Convexification for Hypersonic Reentry Guidance,} , 2024.
\newblock Under review for the AIAA Jounral of Guidance, Control, and Dynamics. {Available at }\url{https://arxiv.org/abs/2411.08361}.

\bibitem[{Mceowen et~al.(2025)Mceowen, Calderone, Tiwary, Zhou, Kim, Elango, and A{\c{c}}ikmese}]{mceowen2025}
Mceowen, S., Calderone, D.~J., Tiwary, A., Zhou, J. S.~K., Kim, T., Elango, P., and A{\c{c}}ikmese, B., \enquote{Auto-tuned Primal-dual Successive Convexification for Hypersonic Reentry Guidance,} \emph{{AIAA Scitech Forum}}, 2025.

\bibitem[{Chari et~al.(2025)Chari, Kamath, Elango, and A{\c{c}}ikmese}]{doi:10.2514/6.2024-1762}
Chari, G.~M., Kamath, A.~G., Elango, P., and A{\c{c}}ikmese, B., \enquote{Fast Monte Carlo Analysis For 6-DoF Powered-Descent Guidance via GPU-Accelerated Sequential Convex Programming,} \emph{{AIAA Scitech Forum}}, 2025.
\newblock \doi{10.2514/6.2024-1762}.

\bibitem[{Echigo et~al.(2024)Echigo, Sheridan, Buckner, and A{\c{c}}{\i}kme{\c{s}}e}]{kazu2024}
Echigo, K., Sheridan, O., Buckner, S., and A{\c{c}}{\i}kme{\c{s}}e, B., \enquote{{Dispersion Sensitive Optimal Control: A Conditional Value-at-Risk-Based Tail Flattening Approach via Sequential Convex Programming},} \emph{{IEEE Transactions on Control Systems Technology}}, Vol.~32, No.~6, 2024, pp. 2468--2475.
\newblock \doi{10.1109/TCST.2024.3427910}.

\bibitem[{Echigo(2024)}]{kazu2024dissertation}
Echigo, K., \enquote{From Theory towards Flight: Convex Optimization based Approaches for Non-convex, Stochastic, and Realistic Aerospace Missions,} Ph.D. thesis, University of Washington, Seattle, WA, 2024.

\bibitem[{Uzun et~al.(2024)Uzun, Elango, Garoche, and A{\c{c}}{\i}kme{\c{s}}e}]{dgmsr}
Uzun, S., Elango, P., Garoche, P.-L., and A{\c{c}}{\i}kme{\c{s}}e, B., \enquote{Optimization with Temporal and Logical Specifications via Generalized Mean-based Smooth Robustness Measures,} \emph{arXiv preprint arXiv:2405.10996}, 2024.
\newblock \urlprefix\url{https://doi.org/10.48550/arXiv.2405.10996}.

\end{thebibliography}

\end{document}